\documentclass[conference]{IEEEtran}
\IEEEoverridecommandlockouts
\usepackage{cite}
\usepackage{graphicx}
\usepackage{amsmath}
\usepackage{times}
\usepackage{latexsym}
\usepackage{bm}
\usepackage{amssymb}
\usepackage[center]{caption2}
\usepackage{array}
\usepackage{fancyhdr}
\usepackage{cite,graphicx,amsmath,amssymb}
\usepackage{psfrag}
\usepackage{multirow}
\usepackage{amsthm}

\ifCLASSINFOpdf

\else

\fi
\usepackage[dvipsnames,usenames]{color}
\usepackage[usenames,dvipsnames]{color}

\newtheorem{theorem}{Theorem}

\newcommand{\tr}{\text{tr}}

\makeatother

\begin{document}
\title{Data-Aided Secure Massive MIMO Transmission with Active
Eavesdropping}

\author{\IEEEauthorblockN{Yongpeng Wu, Chao-Kai Wen, Wen Chen, Shi Jin, Robert Schober,
and  Giuseppe Caire}

\thanks{The work of Y. Wu was supported in part by NSFC No. 61701301.
The work of C.-K. Wen was supported by the Ministry of Science and Technology of Taiwan under Grant MOST 106-2221-E-110-019.
The work of W. Chen is supported by Shanghai STCSM 16JC1402900 and 17510740700,
by National Science and Technology Major Project 2017ZX03001002-005
and 2018ZX03001009-002, by NSF China  61671294, and by Guangxi NSF 2015GXNSFDA139037.
The work of S. Jin was supported in part by the NSFC under Grant 61531011.}

\thanks{Y. Wu and W. Chen are with the Department of Electronic Engineering, Shanghai Jiao Tong University,
Minhang 200240, China (e-mail: yongpeng.wu@sjtu.edu.cn; wenchen@sjtu.edu.cn).}

\thanks{C. K. Wen is with the Institute of Communications Engineering, National Sun Yat-sen University, Kaohsiung 804,
Taiwan (Email: chaokai.wen@mail.nsysu.edu.tw).}

\thanks{S. Jin is with the National Mobile Communications Research Laboratory,
Southeast University, Nanjing, 210096, P. R. China. (Emails: jinshi@seu.edu.cn).}

\thanks{R. Schober is with Institute for Digital Communications, Universit\"{a}t Erlangen-N\"{u}rnberg,
Cauerstrasse 7, D-91058 Erlangen, Germany (Email: schober@lnt.de).}

\thanks{G. Caire is with Institute for Telecommunication Systems, Technical University Berlin, Einsteinufer 25,
10587 Berlin, Germany (Email: caire@tu-berlin.de). }

}

\maketitle

%\vspace{-0.5cm}

\begin{abstract}
In this paper, we study the design of secure communication for time division duplexing
multi-cell multi-user massive multiple-input multiple-output (MIMO) systems with active
eavesdropping. We assume that the eavesdropper actively attacks the uplink
pilot transmission and the uplink data transmission before eavesdropping
the downlink data  transmission phase of the desired users.
We exploit both the received pilots and data signals for uplink channel
estimation.  We show analytically
that when the number of transmit antennas and the length of the data vector both tend to infinity,
the signals of the desired user and the eavesdropper lie in
different eigenspaces of the received signal matrix at the base station
if their signal powers are different.
This finding reveals that  decreasing (instead of increasing)
the desire user's signal power might be an effective approach
to combat a strong active attack
from an eavesdropper. Inspired by this result, we propose a data-aided
secure downlink transmission scheme and derive an asymptotic achievable secrecy
sum-rate expression for the proposed design.
Numerical results indicate that under strong active attacks,
the proposed design
achieves significant secrecy rate gains compared  to
the conventional design employing matched filter precoding and artificial noise
generation.
\end{abstract}

\section{Introduction}
Wireless networks are widely used in civilian and military applications
and have become an indispensable part of our daily lifes. Therefore, security
is a critical issue for future wireless networks.  Conventional security
approaches based on cryptographic techniques have many well-known
weaknesses.
Therefore, new  approaches to security based on information theoretical
concepts, such as the secrecy capacity of the propagation channel, have been
developed and are collectively referred to as physical layer security \cite{Wyner1975BSTJ,Oggier2011TIT,Wu2012TVT,Wu2017TCom}.

Massive MIMO is a promising approach for efficient transmission of massive amounts of information
and is regarded as one of the ``big three" 5G technologies \cite{Andrews2014JSAC}.
Most studies on physical layer security in massive MIMO systems
assume that the eavesdropper is passive and does not attack the communication
process of the systems \cite{Zhu2014TWC,Chen2015TWC,Chen2016TFS,ZhuJ2017TWC}. However, a smart eavesdropper
can perform the pilot contamination attack to jeopardize
the channel estimation process at the base station \cite{Wu2016TIT}.
Due to the channel hardening effect caused by large antenna arrays,
the pilot contamination attack results in a serious secrecy  threat
to time division duplexing (TDD)-based massive MIMO systems \cite{Wu2016TIT}.

The authors of \cite{Im2015TWC} propose a secret key agreement
protocol for single-cell multi-user massive MIMO systems under
the pilot contamination attack. An estimator for the base station (BS)
is designed to evaluate the information leakage.
Then, the BS and the desired users perform secure communication by adjusting the length
of the secrecy key based on the estimated information leakage.
Other works have studied  how to combat the pilot contamination attack.
The authors of \cite{Basciftci2017} investigate
the pilot contamination attack problem for single-cell multi-user massive MIMO systems
over independent and identically distributed (i.i.d.) fading channels.
The eavesdropper is assumed to only know the pilot signal set whose size scales polynomially
with the number of transmit antenna. For each transmission, the desired users randomly select certain pilot signals
from this set, which are unknown to the eavesdropper.
In this case, it is proved that the impact of the pilot contamination
attack can be eliminated as the number of transmit antenna goes to infinity. For the more pessimistic assumption that the
eavesdropper knows the exact pilot signals of the desired users for each transmission,
the secrecy threat caused by
the pilot contamination attack in multi-cell multi-user massive MIMO systems over correlated fading channels is analyzed in \cite{Wu2016TIT}.
Based on this, three transmission strategies for combating the pilot contamination attack
are proposed. However, the designs in \cite{Wu2016TIT} are not able to guarantee a high (or even a non-zero) secrecy rate for
weakly correlated or i.i.d. fading channels under a strong pilot contamination attack.

In this paper, we investigate secure transmission for i.i.d. fading\footnote{For simplicity of
presentation, we assume i.i.d. fading to present the basic idea of data-aided secure massive MIMO
transmission. The results can be extended to the general case of correlated fading channels
 by combining the techniques in \cite{Wu2016TIT} with those in this paper.
This will be considered in extended journal version of this paper.}
TDD multi-cell multi-user massive MIMO systems under a strong active attack.
We assume the system performs first uplink training followed by an uplink data transmission phase
and a downlink data transmission phase.  The eavesdropper jams the uplink training phase
and the uplink data transmission phase and then eavesdrops the downlink data transmission\footnote{}.

We utilize the uplink transmission data to aid
the channel estimation at the BS. Then, based on the estimated channels,
the BS designs precoders for the downlink transmission.

This paper makes the following key contributions:

\begin{enumerate}

\item We prove that when the number of transmit antennas and the amount
transmitted data both approach infinity, the desired users' and the eavesdropper's signals lie in
different eigenspaces of the uplink received signal matrix
due to their power differences.  Our results reveal
that increasing the power gap between the desired users' and the eavesdropper's signals
is beneficial for separating the desired users and the eavesdropper.
This implies that when facing a strong active attack, decreasing (instead of increasing)
the desired users' signal power could be an effective approach to enable secrete
 communication.

\item Inspired by this observation, we propose a joint uplink and downlink data-aided transmission
scheme to combat strong active attacks from an eavesdropper.
Then, we derive an asymptotic achievable secrecy
sum-rate expression for this scheme. The derived
expression indicates that the impact of an active attack on the uplink
transmission can be completely eliminated by the proposed design.

\item Our numerical results reveal that the proposed design achieves
a good secrecy performance under strong active attacks, while the  conventional design employing matched filter precoding and artificial noise
generation (MF-AN) \cite{Wu2016TIT} is not able to guarantee secure communication
in this case.

\end{enumerate}

\emph{Notation:}  Vectors are denoted by lower-case bold-face letters;
matrices are denoted by upper-case bold-face letters. Superscripts $(\cdot)^{T}$, $(\cdot)^{*}$, and $(\cdot)^{H}$
stand for the matrix transpose, conjugate, and conjugate-transpose operations, respectively. We use  ${\tr}({\bf{A}})$ and ${\bf{A}}^{-1}$
to denote the trace and the
inverse of matrix $\bf{A}$, respectively.
 ${\rm{diag}}\left\{\bf{b}\right\}$ denotes a diagonal matrix
with the elements of vector $\bf{b}$ on its main diagonal.
${\rm{Diag}}\left\{\bf{B}\right\}$  denotes a diagonal matrix containing
the diagonal elements of matrix $\mathbf{B}$ on the main diagonal.
The $M \times M$ identity matrix is denoted
by ${\bf{I}}_M$, and the $M \times N$ all-zero matrix and the $N \times 1$ all-zero vector are denoted by $\bf{0}$.
The fields of complex and real numbers are denoted
by $\mathbb{C}$ and $\mathbb{R}$, respectively. $E\left[\cdot\right]$ denotes statistical
expectation. $[\mathbf{A}]_{mn}$ denotes the element in the
$m$th row and $n$th column of matrix $\mathbf{A}$. $[\mathbf{a}]_{m}$ denotes the $m$th entry
of vector $\mathbf{a}$. $\otimes$ denotes the Kronecker product.
$\mathbf{x} \sim \mathcal{CN} \left( {\mathbf{0},{{\bf{R}}_N}} \right)$
denotes a circularly symmetric complex vector
$\mathbf{x} \in {\mathbb{C}^{N \times 1}}$  with zero mean and covariance matrix
${{\bf{R}}_N}$.
${\rm var} (a) $ denotes  the variance of  random variable $a$.
${\left[ x \right]^ + }$ stands for $\max \left\{ {0,x} \right\}$.
$a \gg b$ means that $a$ is much larger than $b$.
%\newpage

\section{Uplink Transmission} \label{sec:multi}
Throughout the paper, we adopt the following transmission protocol.
We assume the uplink transmission phase, composing the uplink
training and the uplink data transmission, which
is followed by a downlink data transmission phase.

We assume the main objective of the
eavesdropper is to eavesdrop the downlink data.
The eavesdropper chooses to attack  the
uplink transmission phase to impair the channel estimation phase at the BS.
The resulting mismatched channel estimation will increase the information leakage in the subsequent downlink transmission.
In the downlink transmission phase, the eavesdropper does not attack
but focuses on eavesdropping the data.

We study a multi-cell multi-user system with $L + 1$ cells.
We assume an $N_t$-antenna BS and $K$ single-antenna users are present in
each cell. The cells are index by $l = \left(0,\ldots,L\right)$,
where cell $l = 0$ is the cell of interest.
We assume an $N_e$-antenna
active eavesdropper\footnote{An $N_e$-antenna eavesdropper is equivalent to
$N_e$ cooperative single-antenna eavesdroppers.}
is located in the cell of interest
and attempts to eavesdrop the data intended for all users in the cell.
The eavesdropper sends pilot signals and artificial noise
to interfere channel estimation and uplink data transmission\footnote{We note that if the eavesdropper only attacks the channel
estimation phase and remains silent during the uplink data transmission, then the impact of this attack can be easily eliminated
with the joint channel estimation and data detection scheme in \cite{Wen2015ICC}. Therefore, a smart eavesdropper will attack the entire uplink transmission.},
respectively. Let $T$ and $\tau$ denote the coherence time of the channel and the length of
the pilot signal, respectively.  Then, for uplink transmission, the received pilot signal matrix
$\mathbf{Y}_p^{m} \in {\mathbb{C}^{N_t \times \tau}} $ and the received data signal matrix $\mathbf{Y}_d^{m} \in {\mathbb{C}^{N_t \times (T - \tau)}}$
at the BS in cell $m$ are given by\footnote{For notation simplicity, we assume the users in each cell use the same transmit power \cite{Zhu2014TWC}.
Following the similar techniques in this paper, the results can be easily extended to the case of different transmit powers of the users in each cell.}
\begin{align}\label{eq:Yp}
{\mathbf{Y}}_{p}^{m} & =\sqrt {{P_{0}}} \sum\limits_{k = 1}^K { {\bf{h}}_{0k}^m{\boldsymbol{\omega }}_{k}^T}  +  \sum\limits_{l = 1}^L {\sum\limits_{k = 1}^K { \sqrt {{P_{l}}} {\bf{h}}_{lk}^m{\boldsymbol{\omega }}_{k}^T} }
\nonumber \\
&   + \sqrt {\frac{{{P_e}}}{{K{N_e}}}} {\bf{{H}}}_{e}^m\sum\limits_{k = 1}^K {{{\bf{W}}_k}}  + {\mathbf{N}_p^m}
\end{align}
\begin{align}\label{eq:Yd}
{\mathbf{Y}}_{d}^{m} & = \sqrt {{P_{0}}} \sum\limits_{k = 1}^K { {\bf{h}}_{0k}^m{\mathbf{d}}_{0k}^T}  +  \sum\limits_{l = 1}^L {\sum\limits_{k = 1}^K { \sqrt {{P_{l}}} {\bf{h}}_{lk}^m {\mathbf{d}}_{lk}^T} }
\nonumber \\
 &  + \sqrt {\frac{{{P_e}}}{{N_e}}} {\bf{{H}}}_{e}^m \mathbf{A}  + {\mathbf{N}_d^m}
\end{align}
where $P_{0}$,  ${{\boldsymbol{\omega}}_{k}} \in \mathbb{C} {^{\tau  \times 1}}$,
and ${\mathbf{d}}_{0k} \sim \mathcal{CN} \left( {\mathbf{0},{{\bf{I}}_{T - \tau}}} \right)$ denote the average transmit power, the pilot sequence,
and the uplink transmission data  of the $k$th user in cell of interest, respectively.
It is assumed that the same $K$ orthogonal pilot sequences
are used in each cell where ${\boldsymbol{\omega }}_{k}^H{{\boldsymbol{\omega }}_{k}} = \tau$
and ${\boldsymbol{\omega }}_{k}^H{{\boldsymbol{\omega }}_{l}} = 0$.
$P_{l}$ and ${\mathbf{d}}_{lk}$ denote the average transmit power and
the uplink transmission data  of the $k$th user in the $l$th cell, respectively.
$\mathbf{h}_{lk}^p \sim \mathcal{CN} \left( {\mathbf{0}, {\beta _{lk}^p}{{\bf{I}}_{{N_t}}}} \right)$
denotes the channel between the $k$th user in the $l$th cell and the BS in the $p$th cell,
where ${\beta _{lk}^p}$ is the corresponding large-scale path loss.
$\mathbf{H}_{e}^l$ and $P_e$ denote the channel between the eavesdropper and the base station in
the $l$th cell and the average transmit power of the eavesdropper, respectively.
We assume the columns of $\mathbf{H}_{e}^l$  are i.i.d. with Gaussian distribution
$\mathcal{CN} \left( {\mathbf{0}, {\beta _{e}^l} {{\bf{I}}_{{N_t}}}} \right)$,
where $\beta _{e}^l$ is the large-scale path loss for the eavesdropper.
For the training phase, the eavesdropper attacks all the users in cell of interest.
Therefore, it uses the attacking pilot sequences  $\sum\nolimits_{k=1}^{K} {\mathbf{W}_k}$ \cite{Basciftci2017},
where $\mathbf{W}_k = \left[{{\boldsymbol{\omega}}_{k}} \cdots {{\boldsymbol{\omega}}_{k}} \right]^T \in \mathbb{C} {^{N_t  \times \tau}}$.
For the uplink data transmission phase, the eavesdropper generates artificial noise $\mathbf{A} \in \mathbb{C} {^{N_t \times T - \tau}}$,
whose elements conform  i.i.d. standard Gaussian distribution.
$\mathbf{N}_p^m \in \mathbb{C} {^{N_t \times \tau}} $ and $ \mathbf{N}_d^m \in \mathbb{C} {^{N_t  \times (T - \tau)}} $ are noise matrices
whose columns are  i.i.d. Gaussian distributed with $\mathcal{CN} \left( {\mathbf{0}, {N_0} {{\bf{I}}_{{N_t}}}} \right)$.

We define $\mathbf{Y}_0 = \left[{\mathbf{Y}}_{p}^{0} \quad  {\mathbf{Y}}_{d}^{0} \right]$ and the
eigenvalue decomposition $\frac{1}{{T{N_t}}}{\mathbf{Y}_0}{{\mathbf{Y}_0^H}} = \left[ {{{\bf{v}}_1},
\cdots ,{{\bf{v}}_{{N_t}}}} \right]{\bf{\Sigma }}{\left[ {{{\bf{v}}_1}, \cdots ,{{\bf{v}}_{{N_t}}}} \right]^H}$,
where the eigenvalues on the main diagonal of  matrix ${\bf{\Sigma }}$ are originated in ascending order.
For the following, we make the important assumption that
due  to the strong active attack and the large-scale path loss difference between the cell of interest and other cells,
 ${P_e}{\beta _e^0}$, ${P_0}\beta _{0k}^0$, and $P_l \beta _{lk}^0$ have the relationship
 ${P_e}{\beta _e^0} \gg {P_0}\beta _{0k}^0 \gg P_l \beta _{lk}^0$.
Let $M = (L + 1)K + {N_e}$ and  vector $\left( {{\theta _1}, \cdots ,{\theta _M}} \right)$ has the same
element as  vector $\left({P_1}\beta _{11}^0, \cdots ,{P_L}\beta _{LK}^0, {P_0}\beta _{01}^0, \cdots ,{P_0}\beta _{0K}^0,{P_e}{\beta _e}, \cdots {P_e}{\beta _e}\right)$
but with the elements originated in ascending order
whose index  $1 \le {i_1} \le {i_2} \cdots  \le {i_K} \le M$ satisfies ${\theta _{{i_k}}} = {P_0}\beta _{0k}^0$, $k = 1,2, \cdots ,K$.
Define $\mathbf{V}_{eq}^0 = \left[ {{{\bf{v}}_{{N_t} - M + {i_1}}},{{\bf{v}}_{{N_t} - M + {i_2}}}, \cdots, {{\bf{v}}_{{N_t} - M + {i_K}}}} \right]$.
Define ${{\bf{H}}_0} = \left[ {{\bf{h}}_{01}^0, \cdots {\bf{h}}_{0K}^0} \right]$ and
${{\bf{H}}_I} = \left[ {{\bf{h}}_{11}^0, \cdots {\bf{h}}_{1K}^0, \cdots ,{\bf{h}}_{L1}^0, \cdots ,{\bf{h}}_{LK}^0} \right]$.
Then, we have the following theorem.

\begin{theorem}\label{theo:channel_estimation}
Let ${\mathbf{Z}_{0p}} =  \frac{1}{{\sqrt {T{N_t}} }} \left(\mathbf{V}_{eq}^{0}\right)^H \mathbf{Y}_{p}^{0} = \left[ {{{\bf{z}}_{0p,1}}, \cdots ,{{\bf{z}}_{0p,K}}} \right]$
and ${{\mathbf{H}}_{eq}^{0}} = \frac{1}{{\sqrt {T{N_t}} }}\left(\mathbf{V}_{eq}^{0}\right)^H{{\bf{H}}_0}  = \left[ { {{\bf{h}}_{eq,01}}, \cdots,  {{\bf{h}}_{eq,0K}}} \right]$.
Then, when $T \rightarrow \infty $ and $N_t \rightarrow \infty$, the
minimum mean square error (MMSE) estimate  ${{\bf{\widehat{h}}}_{eq,0k}}$ of ${{\bf{h}}_{eq,0k}}$
based on ${\mathbf{Z}_{0p}}$ is given by
\begin{align}\label{eq:h_est}
{\widehat {\bf{h}}_{eq,0k}} = \frac{{\sqrt {{P_0}} }}{{{P_0}\tau  + {N_0}}}\left( {\sqrt {{P_0}} \tau {{\bf{h}}_{eq,0k}} + {\mathbf{{n}}_{eq}} } \right)
\end{align}
where $ {\mathbf{{n}}_{eq}} = {{\bf{V}}_{eq}^{0}}{{\bf{{\tilde{n}}}}_{eq}}$ and ${{\bf{{\tilde{n}}}}_{eq}} \sim \mathcal{CN} \left( {0,\tau {N_0}{{\bf{I}}_{{N_t}}}} \right)$.
\begin{proof}
Please refer to Appendix \ref{proof:theo:channel_estimation}.
\end{proof}
\end{theorem}

{\emph{Remark 1:}} The basic intuition behind Theorem \ref{theo:channel_estimation} is that when $T \rightarrow \infty $ and $N_t \rightarrow \infty$,
each channel tends to be an eigenvector of the received signal matrix. As a result, we project the received signal matrix along
the eigenspace which corresponds to the desired users' channel. In this case, the impact of the strong active
attack can be effectively eliminated.

{\emph{Remark 2:}} In Theorem \ref{theo:channel_estimation}, we assume that
the coherence time of the channel is significantly larger than
the symbol duration \cite{Muller2014JSTSP}. This assumption can be justified based on the expression
for the coherence time in \cite[Eq. (1)]{Muller2014JSTSP}. For typical speeds of mobile users and typical symbol duration,
the coherence time can be more than
hundreds symbol durations or even more.

{\emph{Remark 3:}} The simulation results in Section IV indicate that a sufficient
power gap between ${P_0}$ and $P_e$ can  guarantee a good secrecy performance
when the number of transmit antennas and the coherence time of the channel
are large but not infinite. We note that allocating more power to
 the desired users to combat a strong active attack is not needed. In contrast,
the larger gap  between ${P_0}\beta _{0k}^0$ and ${P_e}\beta _{e}^0$
will be beneficial to approach the channel estimation result in Theorem \ref{theo:channel_estimation}.
This implies that  \emph{decreasing} the power of the desire users  can be an effective
secure transmission strategy under a strong active attack.

{\emph{Remark 4:}} We can use large dimension random matrix theory \cite{Couillet2011book} to obtain a more
accurate approximation for the eigenvalue distribution of $\frac{1}{{T{N_t}}}\mathbf{Y} \mathbf{Y}^H$ for the case when
$N_t$ and $T$ are large but not infinite.
Then,  power design policies for $P_0$, $P_l$, and $P_e$ can be obtained.
This will be discussed in the extended journal version of this work.

Based on Theorem \ref{theo:channel_estimation}, we can design the precoders
for downlink transmission.

\section{Downlink Transmission}
In this section, we consider the downlink transmission.
We assume the BSs in all $L + 1$ cells perform  channel
estimation according to Theorem 1 by replacing ${\widehat {\bf{h}}_{eq,0k}}$,
${{\bf{h}}_{eq,0k}}$, $P_0$, and ${{\bf{V}}_{eq}^{0}}$
by ${\widehat {\bf{h}}_{eq,lk}}$, ${{\bf{h}}_{eq,lk}}$, $P_l$, and
${{\bf{V}}_{eq}^{l}}$, respectively. Then, the $l$th BS designs the transmit
signal as follows
\begin{align}\label{eq:xl}
{{\bf{x}}_l} = \sqrt P \sum\limits_{k = 1}^K {{{\bf{t}}_{lk}}{s_{lk}}}, \quad  l = 0, \cdots, L,
\end{align}
where $P$ is the downlink transmission power,
${{\bf{t}}_{lk}} = \left(\mathbf{V}_{eq}^{l}\right)^H\frac{{{{{\bf{\hat h}}}_{eq,lk}}}}{{\left\| {{{{\bf{\hat h}}}_{eq,lk}}} \right\|}}$,
and $s_{lk}$ is the downlink transmitted signal for the $k$th user in the $l$th cell.

For the proposed precoder
design, the base station only needs to know the statistical channel state information of the eavesdropper ${P_e}\beta _{e}^0$ in order
to construct $\mathbf{V}_0$.  This assumption is justified in \cite{Wu2016TIT}.

Because each user in the cell of interest has the risk of being eavesdropped, an achievable ergodic secrecy sum-rate
can be expressed as \cite{Geraci2012TCom}
\begin{align}\label{eq:sum_rate}
R_{\rm sec} = \sum\limits_{k = 1}^K \left[ R_k - C_k^{\rm eve} \right]^{+}
\end{align}
where $R_k$ and  $C_k^{\rm eve}$ denote an achievable ergodic rate between the BS and the $k$th user and the
ergodic capacity between the BS and the eavesdropper seeking to decode the information of the
$k$th user, respectively.

The received signal ${y_{0k}}$ at the $k$th user in the cell of interest is given by
\begin{align}\label{eq:y0k}
{y_{0k}} & = \sum\limits_{l = 0}^L {{{\left( {{\bf{h}}_{lk}^0} \right)}^H}{{\bf{x}}_l}}  + {{n}_d} \nonumber \\
 & = \sqrt P {\left( {{\bf{h}}_{0k}^0} \right)^H} \left(\mathbf{V}_{eq}^{0}\right)^H \frac{{{{{\bf{\hat h}}}_{eq,0k}}}}{{\left\| {{{{\bf{\hat h}}}_{eq,0k}}} \right\|}}{s_{0k}} \nonumber \\
&  + \sqrt P {\left( {{\bf{h}}_{0k}^0} \right)^H}\left(\mathbf{V}_{eq}^{0}\right)^H \sum\limits_{t = 1,t \ne k}^K {\frac{{{{{\bf{\hat h}}}_{eq,0t}}}}{{\left\| {{{{\bf{\hat h}}}_{eq,0t}}} \right\|}}{s_{0t}}}  \nonumber \\
&  + \sqrt P \sum\limits_{l = 1}^L {{{\left( {{\bf{h}}_{lk}^0} \right)}^H}\left(\mathbf{V}_{eq}^{l}\right)^H\sum\limits_{t = 1}^K {\frac{{{{{\bf{\hat h}}}_{eq,lt}}}}{{\left\| {{{{\bf{\hat h}}}_{eq,lt}}} \right\|}}{s_{lt}}} }  + {{n}}_d.
\end{align}
where ${{n}}_d \sim \mathcal{CN} \left( {{0}, N_{0d}} \right)$ is the noise
in the downlink transmission.

We use a lower bound for the achievable ergodic rate $R_k$ as follows \cite{Jose2011TWC}
\begin{align}\label{eq:Rk_lower}
{\bar{R}_k} = \log \left( {1 + {\gamma_k}} \right)
\end{align}
where
\begin{align}\label{eq:gamma_k}
& {\gamma _k} = \nonumber \\
 & \frac{{{{\left| {E\left[ {g_{0k,k}^0} \right]} \right|}^2}}}{{{N_{0d}} + {\mathop{\rm var}} \left( {g_{0k,k}^0} \right) + \sum\limits_{t = 1,t \ne k}^K {E\left[ {{{\left| {g_{0t,k}^0} \right|}^2}} \right] + \sum\limits_{l = 1}^L {\sum\limits_{t = 1}^K {E\left[ {{{\left| {g_{lt,k}^0} \right|}^2}} \right]} } } }}
\end{align}
and $g_{lt,k}^0 = \sqrt P {\left( {{\bf{h}}_{lk}^0} \right)^H}\left(\mathbf{V}_{eq}^{l}\right)^H\frac{{{{{\bf{\hat h}}}_{eq,lt}}}}{{\left\| {{{{\bf{\hat h}}}_{eq,lt}}} \right\|}}$.

For $C_k^{\rm eve}$, we adopt the same pessimistic assumption as in \cite{Wu2016TIT}, i.e.,
we assume that the eavesdropper can eliminate all interference  from intra and inter-cell users to obtain an upper bound of $C_k^{\rm eve}$ as follows
\begin{align}\label{eq:C_eve_k}
C_{k, {\rm upper}}^{\rm eve}
 = E\left[{{\log }_2}\left( 1 + \frac{P}{{{N_0}}}\frac{g_{\rm eve}} {{{{\left\| {{{{\bf{\hat h}}}_{eq,0k}}} \right\|}^2}}}\right) \right]
\end{align}
where
\begin{align}\label{eq:g_eve_k}
g_{\rm eve} = \left({{{{\bf{\hat h}}}_{eq,0k}}}\right)^H  \left(\mathbf{V}_{eq}^{0}\right) {{\left( {{\bf{H}}_e^0} \right)}^H}\left({{\bf{H}}_e^0} \right)\left(\mathbf{V}_{eq}^{0}\right)^H{{{\bf{\hat h}}}_{eq,0k}}.
\end{align}

Based on (\ref{eq:sum_rate}), (\ref{eq:Rk_lower}), and (\ref{eq:C_eve_k}),
we have the following theorem.

\begin{theorem}\label{theo:achievable_rate}
For the considered multi-cell multi-user massive MIMO system,
an asymptotic achievable secrecy sum-rate for the transmit signal design in (\ref{eq:xl})
is given by
\begin{align}\label{eq:sum_rate_lower}
R_{\rm sec,\, ach} \mathop  \to \limits^{{N_t} \to \infty }   \sum\limits_{k = 1}^K \log \left( {1 + {\bar{\gamma}_k}} \right)
\end{align}
where
\begin{align}\label{eq:gamma_k_bar}
{\bar{\gamma}_k} = \frac{{{a_1}}}{{{N_{0d}} + P\left({a_2} - {a_1} \right) + P (K-1) \beta _{0k}^0} + P K \sum\limits_{l = 1}^L {\beta _{lk}^0} }
\end{align}
\begin{align}\label{eq:a1}
{a_1} = \frac{{{P_0}\tau \left( {{P_0}\tau \beta _{0k}^0\left( {{N_t} + K - 1} \right) + K{N_0}} \right)}}{{{{\left( {{P_0}\tau  + {N_0}} \right)}^2}}}
\end{align}
\begin{align}\label{eq:a2}
& {a_2} =  \nonumber \\
& \frac{{{P_0}\tau {{\left( {\beta _{0k}^0{N_t} + \beta _{0k}^0\left( {K - 1} \right)} \right)}^2} + {N_0}\left( {{N_t}\beta _{0k}^0 + 3\left( {K - 1} \right)\beta _{0k}^0} \right)}}{{{P_0}\tau \beta _{0k}^0\left( {{N_t} + K - 1} \right) + {N_0}}}
\end{align}
\begin{proof}
Please refer to Appendix \ref{proof:theo:achievable_rate}.
\end{proof}
\end{theorem}

Theorem \ref{theo:achievable_rate} is a general expression which
is valid  for arbitrary $K$ and $L$. Also, Theorem \ref{theo:achievable_rate}
indicates that when $N_t$ tends to infinity,
the impact of the active attack from the eavesdropper disappears
if the proposed joint uplink and downlink transmission design is adopted.

\section{Numerical Results}
In this section, we present numerical results to examine the proposed design and the obtained analytical results.
We set $L = 3$, $N_t =128$, $\beta_{0k}^0 = 1$, $k = 1,\cdots,K$, $\beta_{lk}^0 = 0.2$, $k = 1,\cdots,K$, $l = 1,\cdots,L$,
and $P_0 =P_1 = ... = P_L$. We define the signal-to-noise ratio (SNR) as ${\rm SNR} = P/N_{0d}$. Also, we define
$\rho = P_E/P_0K$.

\begin{figure}[!t]
\centering
\includegraphics[width=0.35\textwidth]{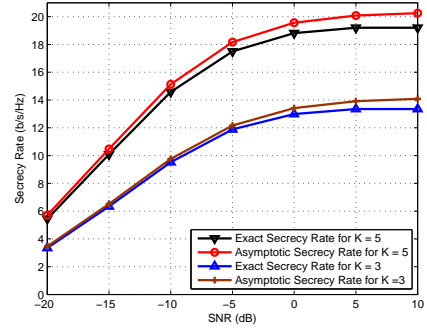}
\caption {\space\space  Secrecy rate vs. SNR for $T = 1024$, $P_0/N_0 =5$ dB, $\rho = 30$, and different numbers of users}
\label{Sec_Asy_SNR_Multi}
\end{figure}

Figure \ref{Sec_Asy_SNR_Multi} plots the asymptotic and exact
secrecy rate performance vs. the
SNR for $T = 1024$, $P_0/N_0 =5$ dB, $\rho = 30$, and different numbers of users, respectively.
The exact secrecy rate is obtained based on Monte Carlo simulation of (\ref{eq:gamma_k}) and (\ref{eq:C_eve_k}).
We note from Figure \ref{Sec_Asy_SNR_Multi} that the asymptotic secrecy rate
in Theorem \ref{theo:achievable_rate} provides a good estimate
for the exact secrecy rate.

\begin{figure}[!t]
\centering
\includegraphics[width=0.35\textwidth]{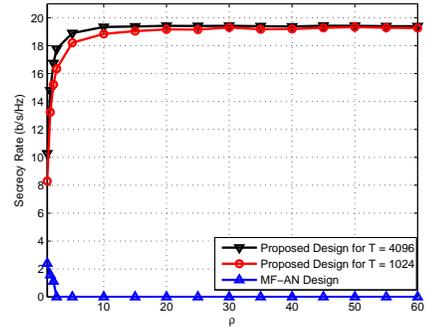}
\caption {\space\space  Secrecy rate vs. $\rho $ for $K = 5$, $P_0/N_0 =5$ dB, SNR = 5dB, and different values of $T$}
\label{Sec_AN_SNR_Multi}
\end{figure}

Figure \ref{Sec_AN_SNR_Multi}
compares the secrecy performance of the proposed design
and the MF-AN design in \cite{Wu2016TIT} for large but finite $N_t$ and $T$
as a function of $\rho$ for $K = 5$, $P_0/N_0 =5$ dB, SNR = 5dB,
and different values of $T$.
We keep ${P_0}$ constant and increase ${P_e}$ to increase $\rho$.
We observe from Figure \ref{Sec_AN_SNR_Multi} that when
the power of the active attack is strong,  the MF-AN design
cannot provide a non-zero secrecy rate. However, our proposed design
performs well in the entire considered range of $\rho$.
As $\rho$ increases, the  gap between ${P_e}{\beta _e}$ and ${P_0}\beta _{0k}^0$ increases as well.
Therefore, the secrecy rate increases with $\rho$ for the proposed design.
 Moreover,  Figure \ref{Sec_AN_SNR_Multi} reveals that increasing $T$ is beneficial for the
secrecy performance of the proposed design.

\section{Conclusions}
In this paper, we have proposed a data-aided
secure transmission scheme for multi-cell  multi-user massive
MIMO systems which are under a strong active attack.
We exploit  the received  uplink  data signal
for  joint uplink channel estimation
and secure downlink transmission.
We show analytically that when the number of transmit antennas and the length of the data vector both approach infinity,
the proposed design can effectively eliminate the impact of an active attack
by an eavesdropper.  Numerical results validate our theoretical analysis and
demonstrate the effectiveness of the proposed design under strong active attacks.

\appendices

\section{Proof of Theorem \ref{theo:channel_estimation}}\label{proof:theo:channel_estimation}
We define $\boldsymbol{\Omega}_0 = \left[{{\boldsymbol{\omega}}_{1}},\cdots, {{\boldsymbol{\omega}}_{K}}\right]^T$,
$\mathbf{D}_{0} =   \sqrt{P_0} \left[{{{\mathbf{d}}}_{01}},\cdots, {{{\mathbf{d}}}_{0K}}\right]^T$, $\boldsymbol{\Omega}_L = \left[\sqrt{P_1} \boldsymbol{\Omega}_0^T,\cdots, \sqrt{P_L} \boldsymbol{\Omega}_0^T \right]^T$,  $\mathbf{D}_{L}  $ \\
$ = \left[ \sqrt{P_1} {{{\mathbf{d}}}_{11}},\cdots, \sqrt{P_1} {{\mathbf{d}}}_{1K}, \cdots,  \sqrt{P_L} {{\mathbf{d}}}_{L1}, \cdots, \sqrt{P_L} {{\mathbf{d}}}_{LK} \right]^T$, ${{\bf{X}}_0} = \left[\sqrt{P_0}  \boldsymbol{\Omega}_0 \quad \mathbf{D}_{0} \right]$,
${{\bf{X}}_I} =  \left[ \boldsymbol{\Omega}_L \quad \mathbf{D}_{L} \right] $, ${{\bf{X}}_e}  = \left[\sqrt {\frac{{{P_E}}}{{K{N_e}}}} \sum\limits_{k = 1}^K {{{\bf{W}}_k}} \quad \sqrt {\frac{{{P_E}}}{{N_e}}}
\mathbf{A} \right]$.

Based on (\ref{eq:Yp}) and (\ref{eq:Yd}), the received signal $\mathbf{Y}_0$ can be re-expressed as
\begin{align}\label{eq:Y0}
{\mathbf{Y}_0}  = {{\bf{H}}_0}{{\bf{X}}_0} + {{\bf{H}}_I}{{\bf{X}}_I} + {{\bf{H}}_e^0}{{\bf{X}}_e} + {\bf{N}}
\end{align}
where ${\bf{N}} = \left[\mathbf{N}_p^0 \quad \mathbf{N}_d^0 \right]$.

When $ T\rightarrow \infty $, based on \cite[Corollary 1]{Evans2000TIT},
we obtain (\ref{YYH-infty3}) given at the top of the next page,
  \begin{figure*}[!ht]
\begin{align}
& \frac{1}{{{N_t}T}}{\bf{Y}}_0{{\bf{Y}}_0^H}\mathop  \to \limits^{ T \to \infty } \frac{1}{{{N_t}T}}{{\bf{H}}_0}{{\bf{X}}_0}{\bf{X}}_0^H{\bf{H}}_0^H + \frac{1}{{{N_t}T}}{{\bf{H}}_I}{{\bf{X}}_I}{\bf{X}}_I^H{\bf{H}}_I^H + \frac{1}{{{N_t}T}}{{\bf{H}}_e^0}{{\bf{X}}_e}{\bf{X}}_e^H \left({\bf{H}}_e^0\right)^H + \frac{{{N_0}}}{{{N_t}}}{{\bf{I}}_{{N_t}}} \nonumber \\
& = \frac{1}{{{N_t}}}\left[ {\begin{array}{*{20}{c}}
{{{\bf{U}}_W}}&{{{\bf{H}}_I}{\bf{B}}_I^{ - 1/2}}&{{{\bf{H}}_e}\beta _e^{ - 1/2}}&{{{\bf{H}}_0}{\bf{B}}_0^{ - 1/2}}
\end{array}} \right] \nonumber \\
& \left[ {\begin{array}{*{20}{c}}
{{N_0}{{\bf{I}}_{{N_t} - M}}}&{}&{}&{}\\
{}&{\frac{{{\bf{B}}_I^{1/2}{{\bf{X}}_I}{\bf{X}}_I^H{\bf{B}}_I^{1/2}}}{T} + {N_0}{{\bf{I}}_{\left( {L - 1} \right)K}}}&{}&{}\\
{}&{}&{\frac{{{\beta _e}{{\bf{X}}_e}{\bf{X}}_e^H}}{T} + {N_0}{{\bf{I}}_{{N_e}}}}&{}\\
{}&{}&{}&{\frac{{{\bf{B}}_0^{1/2}{{\bf{X}}_0}{\bf{X}}_0^H{\bf{B}}_0^{1/2}}}{T} + {N_0}{{\bf{I}}_K}}
\end{array}} \right]
\left[ {\begin{array}{*{20}{c}}
{{\bf{U}}_W^H}\\
{{\bf{B}}_I^{ - 1/2}{\bf{H}}_I^H}\\
{\beta _e^{ - 1/2}{\bf{H}}_e^H}\\
{{\bf{B}}_0^{ - 1/2}{\bf{H}}_0^H}
\end{array}} \right] \nonumber \\
& \mathop \to \limits^{T \to \infty }  \frac{1}{{{N_t}}}  \mathbf{U}_Y
\left[ {\begin{array}{*{20}{c}}
{{N_0}{{\bf{I}}_{{N_t} - M}}}&{}&{}&{} \\
{}&{{\mathbf{P}_I}{{\bf{B}}_I} + {N_0}{{\bf{I}}_{\left( {L - 1} \right)K}}}&{}&{}\\
{}&{}&{\left( {{\beta _e^0}{P_e} + {N_0}} \right){{\bf{I}}_{{N_e}}}}&{}\\
{}&{}&{}&{{P_0}{{\bf{B}}_0} + {N_0}{{\bf{I}}_K}}
\end{array}} \right] \mathbf{U}_Y^H   \label{YYH-infty3}
\end{align}
%\vspace{2cm}
 \hrulefill
 \vspace*{4pt}
% The spacer can be tweaked to stop underfull vboxes.
\end{figure*}
where
\begin{align}
\mathbf{U}_Y & = \left[ {\begin{array}{*{20}{c}}
{{{\bf{U}}_W}}&{{{\bf{H}}_I}{\bf{B}}_I^{ - 1/2}}&{{{\bf{H}}_e}\beta _e^{ - 1/2}}&{{{\bf{H}}_0}{\bf{B}}_0^{ - 1/2}}
\end{array}} \right] \\ \label{YYH-B0}
{{\bf{B}}_0} & = {\rm{diag}} \left( {\beta _{01}^0, \cdots ,\beta _{0K}^0} \right) \\
\label{YYH-BI}
{{\bf{B}}_I} & = {\rm{diag}}\left( {\beta _{11}^0, \cdots ,\beta _{1K}^0, \cdots ,\beta _{L1}^0, \cdots ,\beta _{LK}^0} \right) \\
\label{YYH-BI}
{{\bf{P}}_I} & = {\rm{diag}}\left( {P_1, \cdots ,P_1, \cdots ,P_L, \cdots ,P_L } \right)
\end{align}
and ${{{\bf{U}}_W}} \in \mathbb{C} {^{N_t \times \left(N_t - M \right)}}$ has orthogonal columns.

When $N_t \rightarrow \infty$, we have
\begin{align}\label{eq:UY}
\frac{1}{{{N_t}}}  \mathbf{U}_Y^H \mathbf{U}_Y \mathop \to \limits^{{N_t} \to \infty} \mathbf{I}_{N_t}.
\end{align}

From (\ref{YYH-infty3})--(\ref{eq:UY}), we know that for $T \to \infty$, $N_t \to \infty$,
$\mathbf{U}_Y$ is the right singular matrix
of ${\bf{Y}}_0$. Therefore, we obtain
\begin{align}\label{eq:Z}
& {\bf{Z}} = \frac{1}{{\sqrt {T{N_t}} }}{\left(\mathbf{V}_{eq}^{0}\right)^H}{\bf{Y}}_p^0 \mathop \to \limits^{{N_t} \to \infty } \nonumber  \\
&  \frac{1}{{\sqrt {T{N_t}} }}{\left(\mathbf{V}_{eq}^{0}\right)^H}  \sqrt{P_0}  \boldsymbol{\Omega}_0  {{\bf{X}}_0} + \frac{1}{{\sqrt {T{N_t}} }}{\left(\mathbf{V}_{eq}^{0}\right)^H}{\bf{N}}_p^0.
\end{align}

Define $\mathbf{z} = {\rm vec} \left( {\mathbf{Z}_{0p}} \right) $, where ${\mathbf{Z}_{0p}}$ is defined
in Theorem \ref{theo:channel_estimation}. From (\ref{eq:Z}), we can re-express the equivalent received signal during the pilot transmission
phase as follows
\begin{align}\label{eq:pilot_phase}
{\bf{z}} = \sqrt {{P_0}} \sum\limits_{t = 1}^K {\left( {{{\boldsymbol{\omega }}_t} \otimes {{\bf{I}}_K}} \right){{\bf{h}}_{eq,0t}}}  + {\bf{n}}
\end{align}
where
\begin{align}\label{eq:noise}
{\bf{n}} = \left[ \begin{array}{l}
{\left( {{\bf{V}}_{eq}^0} \right)^H}{{\bf{n}}_{p1}^{0}}\\
 \vdots \\
{\left( {{\bf{V}}_{eq}^0} \right)^H}{{\bf{n}}_{p\tau}^{0} }
\end{array} \right]
\end{align}
and ${\bf{n}}_{pt}^{0}$ in (\ref{eq:noise}) is the $t$th column of  $\mathbf{N}_p^0$.

Based on (\ref{eq:pilot_phase}),  the MMSE estimate of ${{\bf{h}}_{eq,0k}}$ is given by
\begin{align}
 {\widehat {\bf{h}}_{eq,0k}} & = \sqrt {{P_0}} {\left( {{P_0}\tau {{\bf{I}}_K} + {N_0}{{\bf{I}}_K}} \right)^{ - 1}}{\left( {{{\boldsymbol{\omega }}_k} \otimes {{\bf{I}}_K}} \right)^H}{\bf{z}} \nonumber \\
& = \frac{{\sqrt {{P_0}} }}{{{P_0}\tau  + {N_0}}}\left( {\sqrt {{P_0}} \tau {{\bf{h}}_{eq,0k}} + {{\left( {{{\boldsymbol{\omega }}_k} \otimes {{\bf{I}}_K}} \right)}^H}{\bf{n}}} \right). \label{eq:mmse_heq0m_2}
\end{align}
For the noise term in (\ref{eq:mmse_heq0m_2}), we have
\begin{align}
{\left( {{{\boldsymbol{\omega }}_k} \otimes {{\bf{I}}_K}} \right)^H}{\bf{n}} & = \left(\mathbf{V}_{eq}^{0}\right)^H \sum\limits_{t = 1}^\tau  {\omega _{kt}^*} {{\bf{w}}_t} \nonumber  \\
& = \left(\mathbf{V}_{eq}^{0}\right)^H \sum\limits_{t = 1}^\tau  {\omega _{kt}^*} {{\bf{w}}_t} = \left(\mathbf{V}_{eq}^{0}\right)^H {{\bf{\tilde{n}}}_{eq}} \label{eq:mmse_noise}
\end{align}
where ${\omega _{kt}}$ is the $t$th element of ${\boldsymbol{\omega }}_k$.
Combining (\ref{eq:mmse_heq0m_2}) and (\ref{eq:mmse_noise}) completes the proof.

\section{Proof of Theorem \ref{theo:achievable_rate}}\label{proof:theo:achievable_rate}
First, based on the property of MMSE estimates, we know that
$E\left[ {g_{0k,k}^0} \right] = \sqrt P E\left[ {\left\| {{{{\bf{\hat h}}}_{eq,0k}}} \right\|} \right]$.

Based on (\ref{eq:h_est}) and (\ref{YYH-infty3}), we have
\begin{align}
& {\left\| {{{{\bf{\hat h}}}_{eq,0k}}} \right\|^2} = \frac{{{P_0}}}{{{{\left( {{P_0}\tau  + {N_0}} \right)}^2}}} \nonumber \\
& \times {\left( {\sqrt {{P_0}} \tau {{\bf{h}}_{eq,0k}} +  \mathbf{V}_{eq}^{0} {{\bf{\tilde{n}}}_{eq}}} \right)^H}\left( {\sqrt {{P_0}} \tau {{\bf{h}}_{eq,0k}} + \mathbf{V}_{eq}^{0}  {{\bf{\tilde{n}}}_{eq}}} \right) \nonumber
\end{align}
\begin{align}
& = \frac{{{P_0}}}{{{{\left( {{P_0}\tau  + {N_0}} \right)}^2}}} \left[{P_0}{\tau ^2}\frac{1}{{{N_t}}}{\left( {{\bf{h}}_{0k}^0} \right)^H}\left[   {{\bf{h}}_{01}^0, \cdots {\bf{h}}_{0K}^0} \right]{\bf{B}}_0^{ - 1} \right. \nonumber \\
& \times {\left[ {{\bf{h}}_{01}^0, \cdots {\bf{h}}_{0K}^0} \right]^H}{\bf{h}}_{0k}^0  + \frac{1}{{{N_t}}}{\bf{\tilde{n}}}_{eq}^H\left[ {{\bf{h}}_{01}^0, \cdots {\bf{h}}_{0K}^0} \right]{\bf{B}}_0^{ - 1} \nonumber \\
& \times \left. {\left[ {{\bf{h}}_{01}^0, \cdots {\bf{h}}_{0K}^0} \right]^H}{{\bf{\tilde{n}}}_{eq}} \right] \nonumber \\
& = \frac{{{P_0}}}{{{{\left( {{P_0}\tau  + {N_0}} \right)}^2}}}\left[{P_0}{\tau ^2}\frac{1}{{{N_t}}}\sum\limits_{t = 1}^K {{{\left( {{\bf{h}}_{0k}^0} \right)}^H}{{\left( {\beta _{0t}^0} \right)}^{ - 1}}{\bf{h}}_{0t}^0} {\left( {{\bf{h}}_{0t}^0} \right)^H}{\bf{h}}_{0k}^0 \right. \nonumber \\
&  \left. + \frac{1}{{{N_t}}}\sum\limits_{t = 1}^K {{\bf{\tilde{n}}}_{eq}^H{\bf{h}}_{0t}^0} {\left( {\beta _{0t}^0} \right)^{ - 1}}{\left( {{\bf{h}}_{0t}^0} \right)^H}{{\bf{\tilde{n}}}_{eq}} \right].  \label{eq:h_est_norm_2}
\end{align}

When $N_t \rightarrow \infty$, based on \cite[Corollary 1]{Evans2000TIT}, we have
\begin{align} \label{eq:h_est_1}
&\frac{1}{{{N_t}}}{\left( {{\bf{h}}_{0k}^0} \right)^H}{\bf{h}}_{0t}^0{\left( {\beta _{0t}^0} \right)^{ - 1}}{\left( {{\bf{h}}_{0t}^0} \right)^H}{\bf{h}}_{0k}^0 \nonumber \\
& \mathop  \to \limits^{{N_t} \to \infty } \frac{{\beta _{0k}^0{{\left( {\beta _{0t}^0} \right)}^{ - 1}}}}{{{N_t}}} {\rm tr}\left( {{\bf{h}}_{0t}^0{{\left( {{\bf{h}}_{0t}^0} \right)}^H}} \right)\mathop  \to \limits^{{N_t} \to \infty } \beta _{0k}^0
\end{align}
\begin{align} \label{eq:h_est_2}
\frac{1}{{{N_t}}}{\left( {{\bf{h}}_{0k}^0} \right)^H}{\bf{h}}_{0k}^0{\left( {\beta _{0k}^0} \right)^{ - 1}}{\left( {{\bf{h}}_{0k}^0} \right)^H}{\bf{h}}_{0k}^0 \mathop  \to \limits^{{N_t} \to \infty } \beta _{0k}^0{N_t}
\end{align}
\begin{align} \label{eq:h_est_3}
\frac{1}{{{N_t}}}{\bf{\tilde{n}}}_{eq}^H{\bf{h}}_{0t}^0{\left( {\beta _{0t}^0} \right)^{ - 1}}{\left( {{\bf{h}}_{0t}^0} \right)^H}{{\bf{\tilde{n}}}_{eq}}
\mathop  \to \limits^{{N_t} \to \infty} \tau {N_0}.
\end{align}

Substituting (\ref{eq:h_est_1})--(\ref{eq:h_est_3}) into (\ref{eq:h_est_norm_2}), we have
\begin{align} \label{eq:h_est_norm}
& {\left\| {{{{\bf{\hat h}}}_{eq,0k}}} \right\|^2} \mathop  \to \limits^{{N_t} \to \infty} \frac{{{P_0}}}{{{{\left( {{P_0}\tau  + {N_0}} \right)}^2}}}\left( {{P_0}{\tau ^2}\beta _{0k}^0{N_t} + } \right. \nonumber \\
& \left. {{P_0}{\tau ^2}\beta _{0k}^0\left( {K - 1} \right) + K\tau {N_0}} \right).
\end{align}

Next, we evaluate ${\rm var} \left({g_{0k,k}^0}\right)$. First, we obtain
\begin{align}
& {\left( {{\bf{h}}_{0k}^0} \right)^H}\left(\mathbf{V}_{eq}^{0}\right)^H{{{\bf{\hat h}}}_{eq,0k}}{\bf{\hat h}}{_{eq,0k}^H} \mathbf{V}_{eq}^{0}{\bf{h}}_{0k}^0 = \frac{{{P_0}}}{{{{\left( {{P_0}\tau  + {N_0}} \right)}^2}}}{\left( {{\bf{h}}_{0k}^0} \right)^H} \nonumber \\
& \times \frac{1}{{\sqrt {{N_t}} }}\left[ {{\bf{h}}_{01}^0, \cdots {\bf{h}}_{0K}^0} \right]{\bf{B}}_0^{ - 1/2}\left( {\sqrt {{P_0}} \tau {{\bf{h}}_{eq,0k}} + \mathbf{V}_{eq}^{0} {{\bf{\tilde{n}}}_{eq}}} \right) \nonumber \\
& \times  {\left( {\sqrt {{P_0}} \tau {{\bf{h}}_{eq,0k}} + \mathbf{V}_{eq}^{0} {{\bf{\tilde{n}}}_{eq}}} \right)^H}\frac{1}{{\sqrt {{N_t}} }}{\bf{B}}_0^{ - 1/2}{\left[ {{\bf{h}}_{01}^0, \cdots {\bf{h}}_{0K}^0} \right]^H}{\bf{h}}_{0k}^0 \nonumber \\
& \mathop  \to \limits^{{N_t} \to \infty}  \frac{{{P_0}}}{{{{\left( {{P_0}\tau  + {N_0}} \right)}^2}}}{\left( {{\bf{h}}_{0k}^0} \right)^H}\frac{1}{{\sqrt {{N_t}} }}\left[ {{\bf{h}}_{01}^0, \cdots {\bf{h}}_{0K}^0} \right]{\bf{B}}_0^{ - 1/2} \left[\frac{1}{{{N_t}}} \right. \nonumber \\
&\times {P_0} {\tau ^2}{\bf{B}}_0^{ - 1/2} {\left[ {{\bf{h}}_{01}^0, \cdots {\bf{h}}_{0K}^0} \right]^H}\left( {{\bf{h}}_{0k}^0} \right){\left( {{\bf{h}}_{0k}^0} \right)^H}\left[ {{\bf{h}}_{01}^0, \cdots {\bf{h}}_{0K}^0} \right]  \nonumber \\
&  \times{\bf{B}}_0^{ - 1/2} + \frac{1}{{{N_t}}}{\bf{B}}_0^{ - 1/2}{\left[ {{\bf{h}}_{01}^0, \cdots {\bf{h}}_{0K}^0} \right]^H}{{\bf{\tilde{n}}}_{eq}}{\bf{\tilde{n}}}_{eq}^H\left[ {{\bf{h}}_{01}^0, \cdots {\bf{h}}_{0K}^0} \right] \nonumber \\
\end{align}
\begin{align}
& \left. \times {\bf{B}}_0^{ - 1/2}\right] \frac{1}{{\sqrt {{N_t}} }}{\bf{B}}_0^{ - 1/2}{\left[ {{\bf{h}}_{01}^0, \cdots {\bf{h}}_{0K}^0} \right]^H}{\bf{h}}_{0k}^0 \nonumber   \\
& = \frac{{{P_0}}}{{{{\left( {{P_0}\tau  + {N_0}} \right)}^2}}}\left[\frac{{{P_0}{\tau ^2}}}{{N_t^2}}{\left( {{\bf{h}}_{0k}^0} \right)^H}\left[ {{\bf{h}}_{01}^0, \cdots {\bf{h}}_{0K}^0} \right]{\bf{B}}_0^{ - 1} \right. \nonumber \\
& \times {\left[ {{\bf{h}}_{01}^0, \cdots {\bf{h}}_{0K}^0} \right]^H}\left( {{\bf{h}}_{0k}^0} \right){\left( {{\bf{h}}_{0k}^0} \right)^H}\left[ {{\bf{h}}_{01}^0, \cdots {\bf{h}}_{0K}^0} \right]{\bf{B}}_0^{ - 1} \nonumber  \\
& \times {\left[ {{\bf{h}}_{01}^0, \cdots {\bf{h}}_{0K}^0} \right]^H}{\bf{h}}_{0k}^0 + \frac{1}{{N_t^2}}{\left( {{\bf{h}}_{0k}^0} \right)^H}\left[ {{\bf{h}}_{01}^0, \cdots {\bf{h}}_{0K}^0} \right]{\bf{B}}_0^{ - 1} \nonumber \\
& \times {\left[ {{\bf{h}}_{01}^0, \cdots {\bf{h}}_{0K}^0} \right]^H}{{\bf{\tilde{n}}}_{eq}}{\bf{\tilde{n}}}_{eq}^H\left[ {{\bf{h}}_{01}^0, \cdots {\bf{h}}_{0K}^0} \right]{\bf{B}}_0^{ - 1}\nonumber  \\
& \left.\times {\left[ {{\bf{h}}_{01}^0, \cdots {\bf{h}}_{0K}^0} \right]^H}{\bf{h}}_{0k}^0 \right]. \label{eq:h_0k_v0_2}
\end{align}

From (\ref{eq:h_est_1}) and (\ref{eq:h_est_2}), we have
\begin{align} \label{eq:h_0k0k}
& \frac{1}{{{N_t}}}{\left( {{\bf{h}}_{0k}^0} \right)^H}\left[ {{\bf{h}}_{01}^0, \cdots {\bf{h}}_{0K}^0} \right]{\bf{B}}_0^{ - 1}{\left[ {{\bf{h}}_{01}^0, \cdots {\bf{h}}_{0K}^0} \right]^H}\left( {{\bf{h}}_{0k}^0} \right) \nonumber \\
&\mathop  \to \limits^{{N_t} \to \infty } \beta _{0k}^0{N_t} + \beta _{0k}^0\left( {K - 1} \right) \nonumber \\
&\frac{1}{{N_t^2}}{\left( {{\bf{h}}_{0k}^0} \right)^H}\left[ {{\bf{h}}_{01}^0, \cdots {\bf{h}}_{0K}^0} \right]{\bf{B}}_0^{ - 1}{\left[ {{\bf{h}}_{01}^0, \cdots {\bf{h}}_{0K}^0} \right]^H}{{\bf{\tilde{n}}}_{eq}} \nonumber \\
&\times {\bf{\tilde{n}}}_{eq}^H\left[ {{\bf{h}}_{01}^0, \cdots {\bf{h}}_{0K}^0} \right]{\bf{B}}_0^{ - 1}{\left[ {{\bf{h}}_{01}^0, \cdots {\bf{h}}_{0K}^0} \right]^H}{\bf{h}}_{0k}^0 \nonumber \\
\end{align}

Also, we have
\begin{align}
& = \frac{1}{{N_t^2}}{\bf{\tilde{n}}}_{eq}^H\left[ {{\bf{h}}_{01}^0, \cdots {\bf{h}}_{0K}^0} \right]{\bf{B}}_0^{ - 1}{\left[ {{\bf{h}}_{01}^0, \cdots {\bf{h}}_{0K}^0} \right]^H}{\bf{h}}_{0k}^0{\left( {{\bf{h}}_{0k}^0} \right)^H} \nonumber \\
& \times \left[ {{\bf{h}}_{01}^0, \cdots {\bf{h}}_{0K}^0} \right]{\bf{B}}_0^{ - 1}{\left[ {{\bf{h}}_{01}^0, \cdots {\bf{h}}_{0K}^0} \right]^H}{{\bf{\tilde{n}}}_{eq}} \nonumber \\
& \mathop  \to \limits^{{N_t} \to \infty }  \frac{1}{{N_t^2}}\tau {N_0} {\rm tr}\left( {\left[ {{\bf{h}}_{01}^0, \cdots {\bf{h}}_{0K}^0} \right]{\bf{B}}_0^{ - 1}{{\left[ {{\bf{h}}_{01}^0, \cdots {\bf{h}}_{0K}^0} \right]}^H}{\bf{h}}_{0k}^0 } \right. \nonumber \\
& \times \left. {{{\left( {{\bf{h}}_{0k}^0} \right)}^H}\left[ {{\bf{h}}_{01}^0, \cdots {\bf{h}}_{0K}^0} \right]{\bf{B}}_0^{ - 1}{{\left[ {{\bf{h}}_{01}^0, \cdots {\bf{h}}_{0K}^0} \right]}^H}} \right) \nonumber \\
&  = \frac{{\tau {N_0}}}{{N_t^2}}{\left( {{\bf{h}}_{0k}^0} \right)^H}\left[ {{\bf{h}}_{01}^0, \cdots {\bf{h}}_{0K}^0} \right]{\bf{B}}_0^{ - 1}{\left[ {{\bf{h}}_{01}^0, \cdots {\bf{h}}_{0K}^0} \right]^H} \nonumber \\
& \times \left[ {{\bf{h}}_{01}^0, \cdots {\bf{h}}_{0K}^0} \right]{\bf{B}}_0^{ - 1}{\left[ {{\bf{h}}_{01}^0, \cdots {\bf{h}}_{0K}^0} \right]^H}{\bf{h}}_{0k}^0 \nonumber \\
& = \frac{{\tau {N_0}}}{{N_t^2}}{\left( {{\bf{h}}_{0k}^0} \right)^H}\sum\limits_{t = 1}^K {{{\left( {\beta _{0t}^0} \right)}^{ - 1}}{\bf{h}}_{0t}^0} {\left( {{\bf{h}}_{0t}^0} \right)^H} \nonumber
\end{align}
\begin{align}
& \times \sum\limits_{p = 1}^K {{{\left( {\beta _{0p}^0} \right)}^{ - 1}}{\bf{h}}_{0p}^0} {\left( {{\bf{h}}_{0p}^0} \right)^H}{\bf{h}}_{0k}^0  \nonumber \\
& \frac{1}{{N_t^2}}{\left( {{\bf{h}}_{0k}^0} \right)^H}{\left( {\beta _{0t}^0} \right)^{ - 1}}{\bf{h}}_{0t}^0{\left( {{\bf{h}}_{0t}^0} \right)^H}{\left( {\beta _{0p}^0} \right)^{ - 1}}{\left( {{\bf{h}}_{0p}^0} \right)^H}{\bf{h}}_{0p}^0{\bf{h}}_{0k}^0
\end{align}

When $k \ne t \ne p$, we have
\begin{align} \label{eq:h0k0t0p}
& \mathop  \to \limits^{{N_t} \to \infty } \frac{{\beta _{0k}^0}}{{N_t^2}} {\rm tr} \left( {{{\left( {\beta _{0t}^0} \right)}^{ - 1}}{\bf{h}}_{0t}^0{{\left( {{\bf{h}}_{0t}^0} \right)}^H}{{\left( {\beta _{0p}^0} \right)}^{ - 1}}{{\left( {{\bf{h}}_{0p}^0} \right)}^H}{\bf{h}}_{0p}^0} \right) \nonumber \\
& \mathop  \to \limits^{{N_t} \to \infty } \frac{{\beta _{0k}^0}}{{{N_t}}}
\end{align}

When $k = t = p$, we have
\begin{align} \label{eq:h0k0t0p_2}
& \frac{1}{{N_t^2}}{\left( {{\bf{h}}_{0k}^0} \right)^H}{\left( {\beta _{0k}^0} \right)^{ - 1}}{\bf{h}}_{0k}^0{\left( {{\bf{h}}_{0k}^0} \right)^H}{\left( {\beta _{0k}^0} \right)^{ - 1}}{\bf{h}}_{0k}^0{\left( {{\bf{h}}_{0k}^0} \right)^H}{\bf{h}}_{0k}^0 \nonumber  \\
& \mathop \to \limits^{{N_t} \to \infty } {N_t}\beta _{0k}^0
\end{align}

When $k = t \ne p$, $k = p \ne t$, $k \ne t = p$, $k \ne p = t$, we have
\begin{align} \label{eq:h0k0t0p_3}
& \frac{1}{{N_t^2}}{\left( {{\bf{h}}_{0k}^0} \right)^H}{\left( {\beta _{0t}^0} \right)^{ - 1}}{\bf{h}}_{0t}^0{\left( {{\bf{h}}_{0t}^0} \right)^H}{\left( {\beta _{0t}^0} \right)^{ - 1}}{\left( {{\bf{h}}_{0t}^0} \right)^H}{\bf{h}}_{0t}^0{\bf{h}}_{0k}^0 \nonumber \\
& \mathop \to \limits^{{N_t} \to \infty } \beta _{0k}^0
\end{align}

Combining (\ref{eq:h_est_norm}), (\ref{eq:h_0k_v0_2})--(\ref{eq:h0k0t0p_3}), and the definition of ${g_{0k,k}^0}$,
we have
\begin{align} \label{eq:var}
{\rm var} \left({g_{0k,k}^0}\right) = a_2- a_1.
\end{align}
where $a_2$ and $a_1$ are defined in (\ref{eq:a1}) and (\ref{eq:a2}).

For $E\left[ {{{\left| {g_{0t,k}^0} \right|}^2}} \right]$ and $E\left[ {{{\left| {g_{lt,k}^0} \right|}^2}} \right]$, we have
\begin{align} \label{eq:0tk}
E\left[ {{{\left| {g_{0t,k}^0} \right|}^2}} \right] & = PE\left[ {\frac{{{{{\bf{\hat h}}}_{eq,0t}}{\bf{\hat h}}{{_{eq,0t}^H}_{}}_{}}}{{{{\left\| {{{{\bf{\hat h}}}_{eq,0t}}} \right\|}^2}}}{\rm tr}} \left( {{{{\bf{\hat h}}}_{eq,0k}}{\bf{\hat h}}{{_{eq,0k}^H}}} \right) \right] \nonumber \\
& = P\beta _{0k}^0
\end{align}
\begin{align} \label{eq:ltk}
E\left[ {{{\left| {g_{lt,k}^0} \right|}^2}} \right]& = PE\left[ {\frac{{{{{\bf{\hat h}}}_{eq,lt}}{\bf{\hat h}}{{_{eq,lt}^H}_{}}_{}}}{{{{\left\| {{{{\bf{\hat h}}}_{eq,lt}}} \right\|}^2}}}{\rm tr}}\left( {{{{\bf{\hat h}}}_{eq,lk}}{\bf{\hat h}}{{_{eq,lk}^H}}} \right) \right] \nonumber \\
& = P\beta _{lk}^0.
\end{align}

For $C_{k, {\rm upper}}^{\rm eve}$ in (\ref{eq:C_eve_k}), we know from (\ref{YYH-infty3})
that when $N_t \rightarrow \infty$, $ \left(\mathbf{V}_{eq}^{0}\right) \left(\mathbf{H}_e^0\right)^H \rightarrow 0$.
Therefore, we have
\begin{align} \label{c_kupper}
C_{k, {\rm upper}}^{\rm eve} \mathop \to \limits^{{N_t} \to \infty} 0 .
\end{align}

Substituting (\ref{eq:h_est_norm}), (\ref{eq:var}), (\ref{eq:0tk}), (\ref{eq:ltk}), and (\ref{c_kupper})
into (\ref{eq:sum_rate}) completes the proof.

%\bibliographystyle{IEEETran}
%\bibliography{IEEEabrv,Reference}


\end{document}